\documentstyle[aps,epsf]{revtex}
\begin{document}
\draft
\twocolumn[\hsize\textwidth\columnwidth\hsize\csname @twocolumnfalse\endcsname

\title{Fast algorithm for calculating two-photon absorption spectra}
\author{Yoshiyuki Kurokawa$^{1,2}$, Shintaro Nomura$^2$,
Tadashi Takemori$^1$ and Yoshinobu Aoyagi$^2$}
\address{$^1$Institute of Applied Physics, University of Tsukuba,
Tukuba 305-8573, Japan}
\address{$^2$RIKEN, 2-1 Hirosawa, Wako, Saitama 351-0198, Japan}
\date{\today}
\maketitle
\begin{abstract}
We report a numerical calculation of the two-photon absorption 
coefficient of electrons in a binding potential using
the real-time real-space higher-order difference method.
By introducing random vector averaging
for the intermediate state, 
the task of evaluating the 
two-dimensional time integral is reduced to 
calculating two one-dimensional integrals.
This allows the reduction of the computation load 
down to the same order as that for the linear response function.
The relative advantage of the method compared to the straightforward
multi-dimensional time integration is greater for 
the calculation of non-linear response functions of higher order 
at higher energy resolution.

\end{abstract}

]


The measurement of two-photon absorption 
coefficient\cite{PhotonandAtoms} yields different
information from the single-photon absorption measurement, 
since the physical processes involved and
selection rules are different.
Because of this, there has always been a lot of interest in
two-photon absorption of various molecules, crystals and solids
\cite{Bryant1980,Vaid1981,Murayama1995,Li1998,Cotter1992,Oak1993,Tommasi1993,Banfi1994}.
To compare against the experimental data, one would like to 
have theorerical calculations based on some realistic model 
of the material.
However, since the two-photon absorption is a typical non-linear optical
process, its realistic modeling has always proved
difficult for large complex systems
\cite{Bryant1980,Vaid1981,Murayama1995,Li1998,Cotter1992}.

A powerful method that has come to be used widely for large
quantum systems is the real-time real-space higher-order difference 
method\cite{OrderN,HedSi}, in which 
the real-space is represented by discrete mesh points,
and the time development of a system is solved by numerically 
integrating the Schr\"odinger equation for discrete
time steps.  The energy levels and energy eigenstates 
are obtained by Fourier analyzing the numerical solution.
The memory requirement scales
linearly with the number of basis states $N$ compared to 
$N^2$ for matrix diagonalization, 
and the method has proved effective in solving large quantum
systems that cannot be solved by conventional
methods\cite{HedSi,oldHedSi}.
So far, the large computation load has meant that
the actual application of the method has been made
primarily to the calculation of 
linear response functions of one-particle systems
\cite{OrderN,HedSi}.  Nevertheless, 
the potential scale advantage of the 
method when applied to large systems invites the speculation 
that the development of the method will be essential 
in making the calculation of non-linear response functions
of complex many-body quantum systems feasible.

In this article, we report a trial application of the
method to the calculation of the two-photon absorption 
coefficient of non-interacting electrons trapped by a 
binding potential
and exposed to monochromatic light of frequency $\omega$.  

The two-photon absorption coefficient is given by\cite{PhotonandAtoms}
\begin{eqnarray}
&\alpha^{(2)}(\omega)&
=\left(\frac{e}{\hbar}\right)^4
\sum_{E_f > E_F}\sum_{E_i < E_F}\nonumber\\
&&\left| \frac{1}{V}\int_{0}^{\infty}dt_1 \int_{0}^{t_1}dt_2
{\rm e}^{-i(\omega-i\eta)(t_1+t_2)}
\right. \nonumber\\
&\times &
\left. \langle E_f|
{\rm e}^{iHt_1/\hbar}{\bf r}{\rm e}^{-iH(t_1-t_2)/\hbar}
{\bf r}{\rm e}^{-iHt_2/\hbar}
| E_i \rangle
\right|^2
\label{eq:tpa}
\end{eqnarray}
where $H$ is the unperturbed hamiltonian of the system,
${\bf r}$ is the electron coordinate operator,
$\eta$ is the frequency resolution,
$V$ is the volume of the system,
the summation with respect to the initial state $|E_i \rangle$
must be taken over all states below the Fermi level $E_F$, and
the summation with respect to $|E_f \rangle$ over 
the states $E_f > E_F$.

An important ingredient of the real-time real-space 
higher-order difference method is the use of random 
vectors as probes to scan the Hilbert space
\cite{OrderN,HedSi,Skilling1989,Drabold1993,Sankey1994,Silver1994}.
Among various types of random vector is the uniform
amplitude random phase vector
\begin{equation}
|\Phi\rangle=\sum_{n=1}^{N}|n\rangle {\rm e}^{i\phi_n},
\label{eq:randvec}
\end{equation}
whose effectiveness has been amply 
demonstrated\cite{OrderN,HedSi,Skilling1989,Drabold1993,Sankey1994,Silver1994}.
Here, the phases $\phi_n$ are independent 
random variables with uniform distribution in the
range $[-\pi, \pi)$, and 
$|n\rangle~ (n=1,N)$ are the
orthonormal basis states which are localized 
at the mesh points in the real space.
Using the property of completeness
\begin{equation}
\left\langle |\Phi\rangle\langle\Phi| \right\rangle_{\Phi} = I
~~( \mbox{identity operator} )
\label{eq:comp}
\end{equation}
which obtains after averaging over random realizations of 
$|\Phi\rangle$ denoted above by the brackets 
$\left\langle\cdots\right\rangle_{\Phi}$,
the two-photon absorption coefficient
eq.(\ref{eq:tpa}) can be rewritten as
\begin{eqnarray}
\alpha^{(2)}(\omega)
&=&\left(\frac{e}{\hbar}\right)^4\left\langle\left|
\frac{1}{V}    \int_{0}^{\infty}dt_1 \int_{0}^{t_1}dt_2
{\rm e}^{-i(\omega-i\eta)(t_1+t_2)}
\right.\right. \nonumber\\
&\times& 
\langle \Phi|\theta(H-E_F)
{\rm e}^{iHt_1/\hbar}{\bf r}{\rm e}^{-iH(t_1-t_2)/\hbar}
\nonumber\\
&\times& \left. \left.
{\bf r}{\rm e}^{-iHt_2/\hbar}
\theta(E_F-H)|\Phi \rangle 
\right|^2
\right\rangle_{\Phi}.
\label{eq:tpa-1}
\end{eqnarray}
The operator step function\cite{Sankey1994}
\begin{equation}
\theta(X)=\sum_{X_i}|X_i\rangle \theta(X_i) \langle X_i|
\end{equation}
can be explicitly constructed for any bounded 
hermitian operator $X$ without solving for the 
eigenvalues $X_i$ and eigenvectors $|X_i\rangle$.  
In our calculation, 
we used an algorithm based on Chebyshev
polynomial expansion, which yields $\theta(X)$ as a
polynomial of the operator $X$
\cite{OrderN,HedSi,Sankey1994,Silver1994,Silver1996,Numerical}.

If eq.(\ref{eq:tpa-1}) were to be numerically 
implemented straightforwardly, the matrix element
inside the integral would have to be obtained for all necessary
combinations of time variables $t_1$ and $t_2$.
One would then start with a random vector $| \Phi \rangle$, 
solve the time development according to the Schr\"odinger
equation, mutiplying operators ($\bf r$'s and $\theta$'s)
along the way as required, and take the inner product 
with $| \Phi \rangle$ at the end.  
The number of discrete time steps required for a calculation 
of energy resolution $\eta$ scales as $\eta^{-1}$,
so that the direct implementation of eq.(\ref{eq:tpa-1})
requires a computation load that grows as $\eta^{-2}$.
On top of it, the calculation has to be repeated for a 
number of different realizations of $| \Phi \rangle$ for
random averaging.  This is necessary to reduce
the fluctuation arising from the use of random vectors, whose 
amplitude can be of the same order in magnitude 
as the final result itself.  The scale of such 
computation can easily overwhelm the capacity of 
any computing facility in existence.

However, the computational load can be greatly reduced 
by inserting the completeness relation eq.(\ref{eq:comp}) 
in the matrix element of eq.(\ref{eq:tpa-1}) to decompose
it into two factors.
The absorption coefficient is then given by
\begin{eqnarray}
&&\alpha^{(2)}_{2}(\omega)\nonumber\\
&=&\left(\frac{e}{\hbar}\right)^4\left\langle\left|
\frac{1}{V}    \int_{0}^{\infty}dt_1 \int_{0}^{t_1}dt_2
{\rm e}^{-i(\omega-i\eta)(t_1+t_2)}
\right.\right. \nonumber\\
&\times& 
\left\langle \langle \Phi|
\theta(H - E_F)
{\rm e}^{iHt_1/\hbar}{\bf r}{\rm e}^{-iHt_1/\hbar}
\theta(E_c - H)
|\Phi' \rangle 
\right. \nonumber\\
&\times& \left. \left.
\left. \langle \Phi'|
{\rm e}^{iHt_2/\hbar}{\bf r}{\rm e}^{-iHt_2/\hbar}
\theta(E_F - H)
|\Phi \rangle \right\rangle_{\Phi'}
\right|^2 \right\rangle_{\Phi},
\label{eq:tpa-2}
\end{eqnarray}
where $|\Phi \rangle$ and $|\Phi' \rangle$ are 
mutually independent random vectors,
and $E_c$ is the cutoff energy to be explained below.
The most costly process of integrating the Schr\"odinger equation 
is now used only to obtain two complex functions
of a single time variable instead of a bivariate function
with two time variables.  
Once the necessary complex-valued functions 
have been calculated and stored, the two-dimensional time 
integration may easily be done with a small computer.
The computational load therefore scales only as $\eta^{-1}$  
with the energy resolution.

The benefit must be weighed against the increased
cost of having to average over random realizations of
intermediate states $|\Phi' \rangle$.
The statistical variance arising from the random sampling
is independent of the number of time steps used in the calculation,
but is controlled only by the number of random samples taken.
Therefore, the relative advantage of the use of eq.(\ref{eq:tpa-2})
over the straightforward integration of eq.(\ref{eq:tpa-1}) increases
as higher energy resolution is required.
For the actual numerical implementation of the random sampling,
it is essential that one has control over
the extent of the Hilbert space to be probed.
In eq.(\ref{eq:tpa-2}), the extra cutoff factor 
$\theta(E_c - H)$ is inserted for this purpose.
The final result should be independent of the 
cutoff energy $E_c$ if it is taken sufficiently large.
However, a large value of $E_c$ entails large random 
fluctuation, so that larger number of samples will be required
to average out the noise to achieve the same accuracy.
Therefore, the value of $E_c$ must be set as small as 
possible provided that its interference with the
final result is kept within the margin of tolerance.

In order to compare the CPU time for the two methods, 
we have computed eqs.(\ref{eq:tpa-1}) and (\ref{eq:tpa-2})
for the case of a parabolic binding potential 
\begin{equation}
V(r)=\frac{m \omega_0^2}{2}r^2 
\label{eq:potential}
\end{equation}
with $\omega_0 = 0.3$ atomic unit (a.u.) and $m$ being the electron mass.
The real space was represented by $16^3$
mesh points to cover a cubic volume of linear dimension $16$ a.u.
For the time development, discrete time steps with
$\Delta t = 0.05$ a.u.
were used for integration of the Schr\"odinger equation.
The Fermi energy was set at $E_F = 3\hbar\omega_0$
and the frequency resolution was $\eta = 8 \times 10^{-2}$ a.u.

\begin{figure}[htb]
\epsfxsize=8.5cm \epsfbox{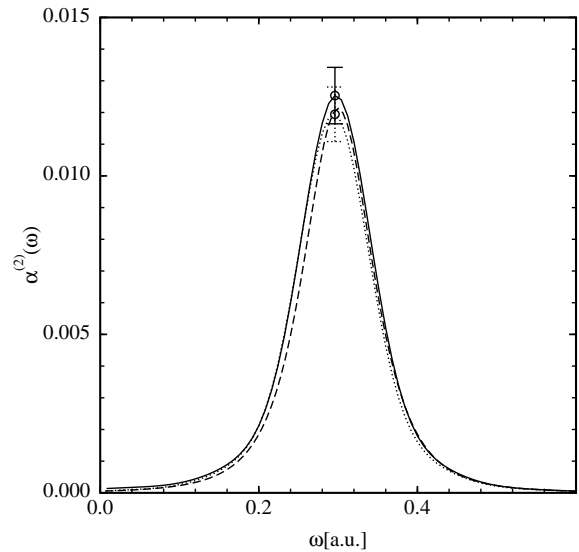}
\caption{
The two-photon absorption coefficient $\alpha^{(2)}(\omega)$ 
of electrons trapped in a parabolic potential.
The results obtained by the use of eqs.(4) and (6)
are shown by the dotted curve and the solid curve, respectively.
The analytical result is shown by the dashed line.}
\label{fig1}
\end{figure}
In fig.\ref{fig1} we compare the results of 
eqs.(\ref{eq:tpa-1}) and (\ref{eq:tpa-2})
with the analytical result for electrons in the parabolic 
potential.  Both numerical results well reproduce the
analytical curve with the standard deviation standing at
7\% at the peak ($\omega_0 = 0.3$ a.u.) for both cases.

The result for the straightforward implementation of 
eq.(\ref{eq:tpa-1}) is an average over 200 runs with 
different $| \Phi \rangle$'s.  The CPU time on a single
processing unit of 
Fujitsu VPP500 was $1.5\times10^{3}$ seconds for each run, 
totaling $3.0\times10^{5}$ seconds ($\simeq$ 83 hours)
to achieve the 7\% accuracy.  For the calculation
according to eq.(\ref{eq:tpa-2}), an average was first taken
over 50 different samples of $|\Phi'\rangle$ with a fixed
$|\Phi \rangle$.  The result was then averaged over
100 different samples of $|\Phi \rangle$.  The total of
5000 runs of integrating the Schr\"odinger equation
took $6.8\times10^{4}$ seconds ($\simeq$ 19 hours),
to achieve the same 7\% accuracy at
the peak.  If the statistical standard deviation is to be brought
down to 1\%, an average over $7^2=49$ times more samples
will have to be taken, which translates to $1.5\times10^{7}$ seconds 
($\simeq$ 174 days)
and $3.3\times10^{6}$ seconds
($\simeq$ 38 days)
for eqs.(\ref{eq:tpa-1}) and (\ref{eq:tpa-2}),
respectively.  In fig.2, we show the estimated CPU time
to achieve 1\% statistical accuracy for various values 
of frequency resolution.  The relative advantage of the 
use of random vectors for the intermediate state should
grow as higher frequency resolution is required.
The actual lapse-time of computation can be reduced 
nearly by an order of magnitude 
if the computation is parallelized to use
all the 30 processing units on 
Fujitsu VPP500 at RIKEN.
\begin{figure}[htb]
\epsfxsize=8.5cm \epsfbox{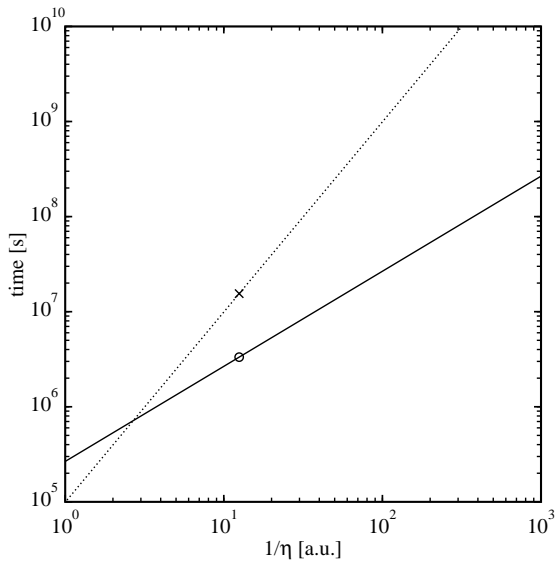}
\label{fig2}
\caption{
The relation between the energy resolution and the total CPU time for
calculating the two-photon absorption coefficient to 1\% statistical
accuracy.
The solid line and the dotted line are for calculation
according to eqs.(6) and (4)
of the text, respectively.
The cross and the circle are the projected CPU time for
energy resolution $\eta = 8\times10^{-2}$ a.u.
based on the actual calculation
performed with the same resolution but to 7\% accuracy.
}
\end{figure}

The present calculation was done for the specific case of
non-interacting electrons trapped in a parabolic potential,
only in order to test the relative advantage of the
algorithm and to compare against the analytical result.  
The computer code admits an arbitrary potential, and work 
is under way to extend the present calculation to electrons
in various pseudo-potentials.  Extension of the algorithm
itself to deal with an interacting $M$-particle system
is straightforward, although solution of the Schr\"odinger
equation in the $3M$-dimensional space soon becomes unmanageable
with increasing $M$.  Of more significance is the potential of 
the present strategy to be used to facilitate the calculation
of nonlinear response functions of higher order.  
A numerical calculation 
of a non-linear response function of order $n$
typically involves computing an $n$-dimensional integral of the
type
\begin{eqnarray}
\int_{0}^{\infty}dt_1
\int_{0}^{t_1}dt_2
\cdots
\int_{0}^{t_{n-1}}dt_n
{\rm e}^{i(\omega_1 t_1 + \omega_2 t_2 + \cdots + \omega_n t_n)}
\nonumber\\
\left\langle \langle \Phi |
\hat{O}_f{\rm e}^{iHt_1/\hbar}
\hat{O}_1{\rm e}^{-iH(t_1-t_2)/\hbar}
\hat{O}_2{\rm e}^{-iH(t_2-t_3)/\hbar}
\right. \nonumber\\
\left. \cdots~
{\rm e}^{-iH(t_{n-1}-t_n)/\hbar}
\hat{O}_{n}{\rm e}^{-iHt_n/\hbar}
\hat{O}_i
| \Phi \rangle \right\rangle_{\Phi},
\label{eq:mult-int}
\end{eqnarray}
where each $\hat{O}$ represents some operator.
While the straightforward evaluation requires computation time
proportional to $\eta^{-n}$, it may be possible to reduce the 
CPU time as far down as to $n$ times that for linear response function 
by inserting the completeness relation eq.(\ref{eq:comp}) and 
decomposing the matrix element into $n$ factors.  
This may be
regarded as a version of quantum Monte Carlo method, and the
use of importance sampling techniques\cite{Drabold1993,Suzuki1986}
will be vital in reducing the evaluation time. 
In fact, the use of $\theta(E_c - H)$ in eq.(\ref{eq:tpa-2}) is one
rudimentary method of improving the sampling efficiency.

The insertion of random intermediate vectors is not the only way
to reduce the CPU time of evaluation of the n-dimensional
integral.
For example, the standard Monte Carlo random sampling technique 
may be applied directly to the evaluation of the 
integral eq.(\ref{eq:mult-int}).
If one adopts uniform random sampling in the n-dimensional time 
space, the CPU time for integration of the Schr\"odinger equation
scales as $\eta^{-1}$ times the number of samples required for
statistical averaging, which is precisely the same as for the
case of random vector insertion above.
In either case, the statistical method of evaluation suffers 
from the familiar negative sign problem\cite{Suzuki1986}, so that
some scheme needs to be devised to improve the sampling efficiency.
Nevertheless, it remains true that 
the size of statistical variance is independent 
of the energy resolution, so that it should be advantageous
to employ statistical methods for the calculation of
high-order nonlinear coefficient at high resolution of frequency.

All calculations reported here were performed on a 
Fujitsu VPP500 at RIKEN.

\end{document}